\documentclass[iop]{emulateapj}
\shortauthors{Ben-Jaffel and Ballester}
\shorttitle{TRANSIT OF EXOMOON PLASMA TORI}
\def \lya\ {Lyman-$\alpha\, $} 
\def \rp {$R_p$}
\def \hdn {HD\,189733}
\begin{document}
\title{TRANSIT OF EXOMOON PLASMA TORI: NEW DIAGNOSIS}
\author{Lotfi Ben-Jaffel \altaffilmark{1}}
\affil{UPMC, University of Paris 06, UMR7095, Institut d'Astrophysique de Paris, 
F-75014, Paris, France; bjaffel@iap.fr}
\altaffiltext{1}{Also at CNRS, UMR7095, Institut d'Astrophysique de Paris, F-75014, Paris, France}
\author{Gilda E. Ballester}
\affil{Lunar and Planetary Laboratory, University of Arizona, 1541 E. University Blvd, Tucson, AZ 85721-0063, USA; gilda@pirl.lpl.arizona.edu}
\begin{abstract}
In the solar system, moons largely exceed planets in number. The {\it Kepler} database has been  shown to be sensitive to exomoon detection down to the mass of Mars, but the first search has been unsuccessful. Here, we use a particles-in-cell code to predict the transit of the plasma torus produced by a satellite. Despite the small size of a moon, the spatial extent of its plasma torus can be large enough to produce substantial transit absorptions. The model is used for the interpretation of {\it Hubble Space Telescope} early ingress absorptions apparently observed during WASP-12\,b and \hdn\,b  UV transits for which no consistent explanation exists. For \hdn\,b an exomoon transiting $\sim 16$\,$R_p$ ahead of the planet and loading $\sim 10^{29}$ C\,II ions/s into space is required to explain the tentative early ingress absorption observed for C\,II. For WASP-12b, a moon transiting $\sim 6$\,$R_p$ ahead from the planet and ejecting $\sim 10^{28}$ Mg\,II ions per second is required to explain the NUV early ingress absorption feature. Interestingly, both \hdn b and WASP-12b predicted satellites are outside the Hill sphere of their planets, an indication that the moons, if present, were not formed in situ but probably captured later. Finally, our simulations show a strong electromagnetic coupling between the  polar regions of planets and the orbital position of the moons, an expected outcome of the unipolar induction DC circuit model. Future observations should test our predictions with a potential opportunity to unambiguously detect the first exomoon plasma torus.
\end{abstract}

\received{March 12, 2014}
\accepted{March 21, 2014}

\keywords{planets and satellites: atmospheres --- planets and satellites: composition --- planets and satellites: detection --- plasmas --- opacity--- ultraviolet: planetary systems}
\section{INTRODUCTION} 
Moons represent a fundamental component of our solar system because they are tightly related to its formation and evolution process. Regular moons represent a sub-population that was formed in situ and evolved with their parent planets. A second sub-population could have been captured after the formation of the planet, leading to a system with an alien body that may have a distinct composition \citep{hua83}. As seen from the planet, a group of satellites represents a complex gravitational system that has several dynamical properties comparable to a planetary system. Therefore, detecting satellites can enhance our understanding of planetary system evolution by extending the spectrum of gravitational configurations of {\it N}-body systems.

\begin{figure}
\centering
\includegraphics[width=6cm,angle=90]{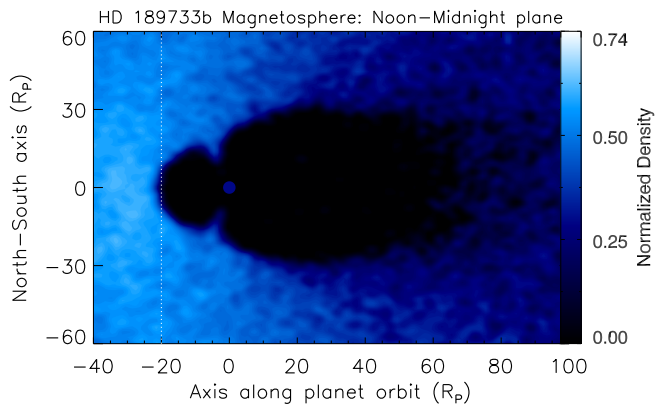}
\caption{{ Three-dimentional PIC simulation of \hdn\,b (or WASP-12b) magnetosphere under sub-Alfv\'enic stellar flow conditions \citep{coh11}. The planet position (0,0) is indicated by a small blue disk. The magnetopause position is also shown. The simulation was obtained for a stellar wind that has a relative speed of $230$\,km/s, an Alfv\'en speed of $\sim 300$\,km/s corresponding to an interplanetary magnetic field of $\sim 4$\,mG, and a sound speed of $\sim 190$\,km/s leading to a sonic Mach number $M_s\sim 1.2$ and Alfv\'en Mach number of $M_v\sim 0.8$. Because C\,II and Mg\,II abundances are negligible in the stellar wind, this structure has no signature in the corresponding transit light curves. Other relevant parameters are discussed in the text.} }
\end{figure}

The large number of satellites in our solar system and the promise of the increasing number of exoplanets being discovered adds to the likelihood of detection of an extrasolar satellite in the near future. Several techniques have been proposed to detect exomoons principally based on the transit technique  \citep{sar99,kip11,sat09,sim12}. Unfortunately, both the small size and mass of moons make their detection a very difficult task \citep{kip11}. Apart from technological limitations, one potential explanation for the non-detection of exomoons could be that for most detected exoplanets orbiting very close to their host stars, the size of their Hill sphere is shrunk significantly, probably limiting their satellite population mostly to bodies captured after the exoplanet formation \citep{dom06,nam10,wei10}.

\begin{figure*}
\centering
\hspace*{-0.4in}
\includegraphics[width=9cm]{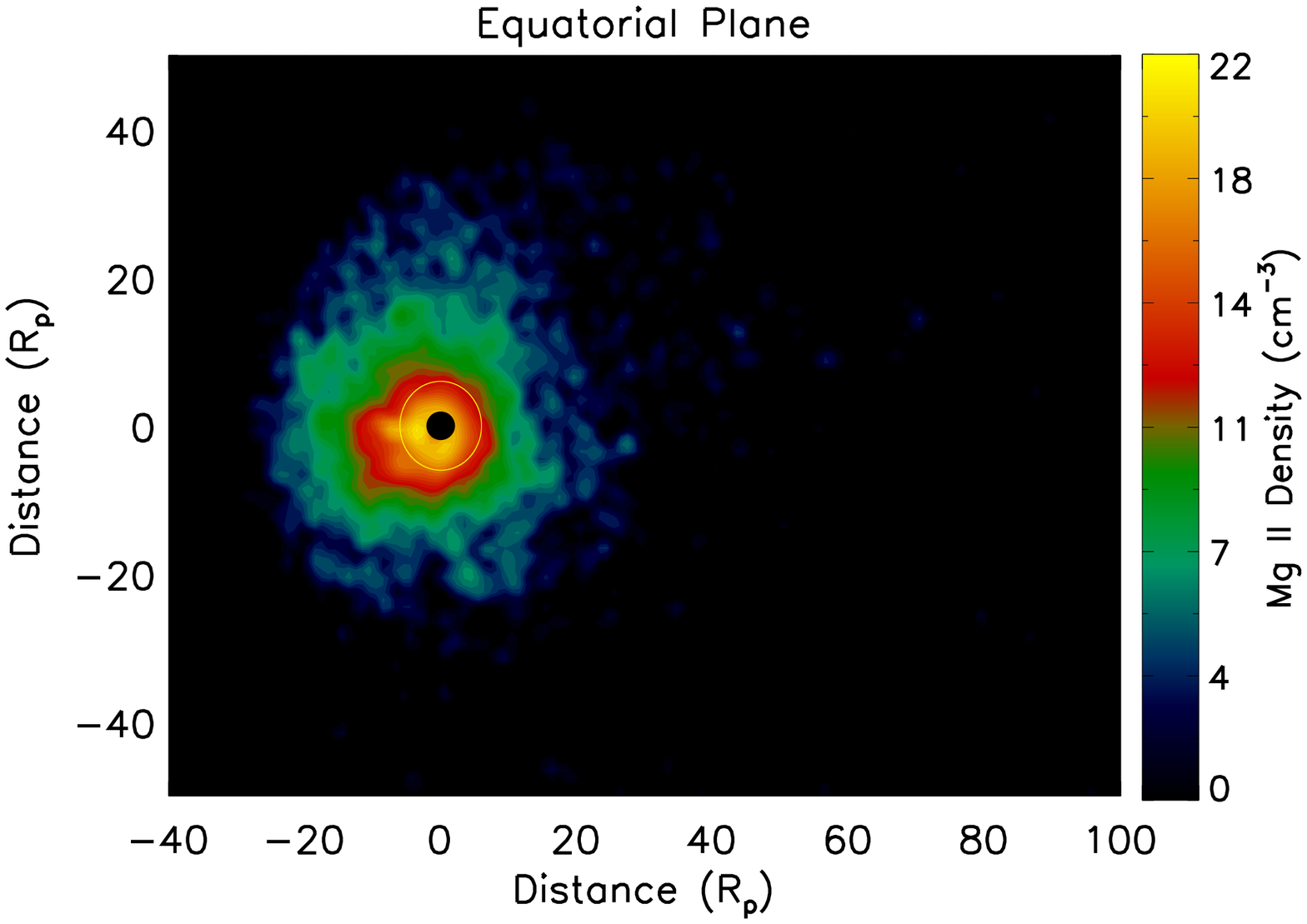}
\includegraphics[width=9cm]{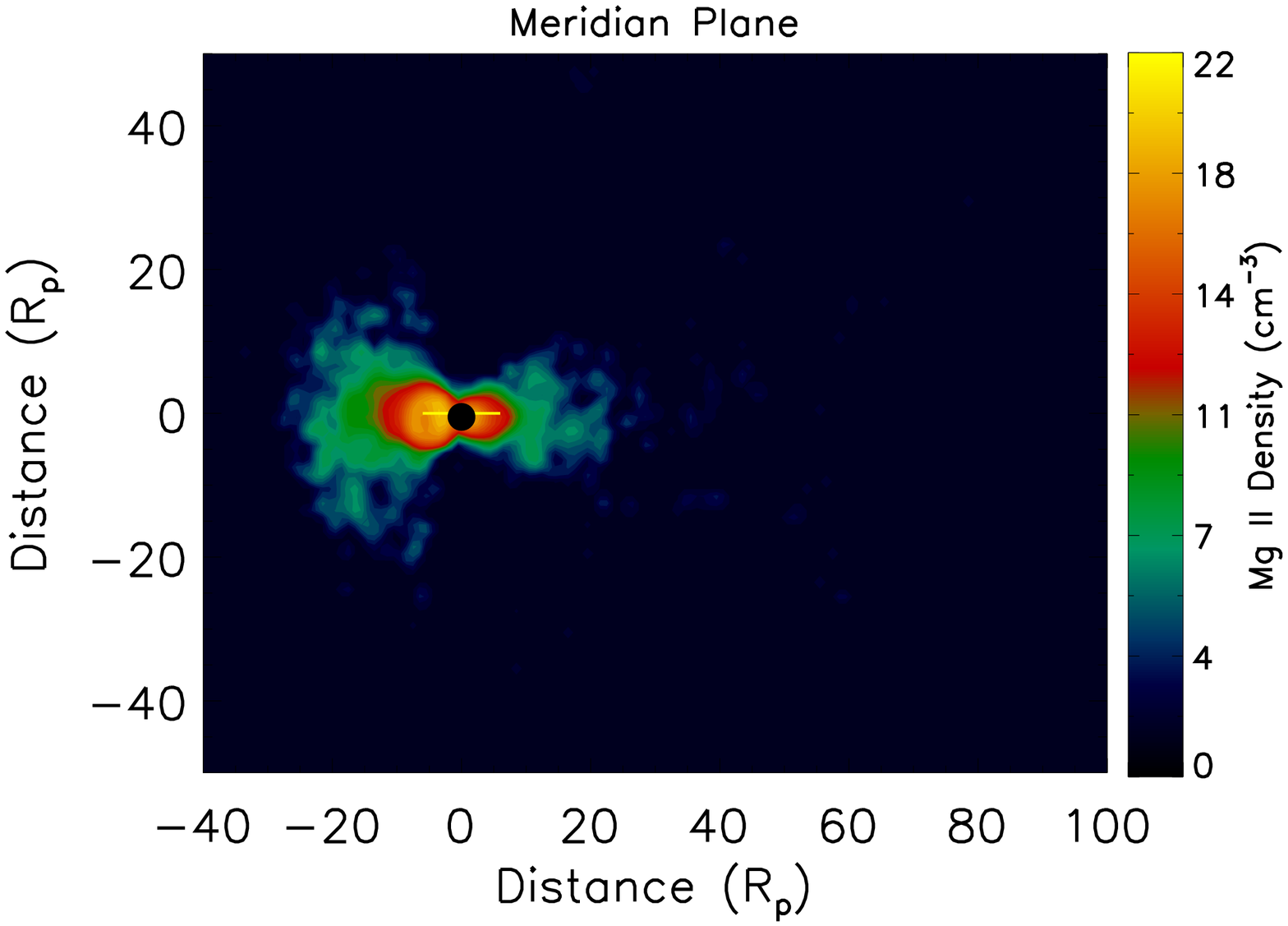}
\hspace*{-0.4in}
\includegraphics[width=9cm]{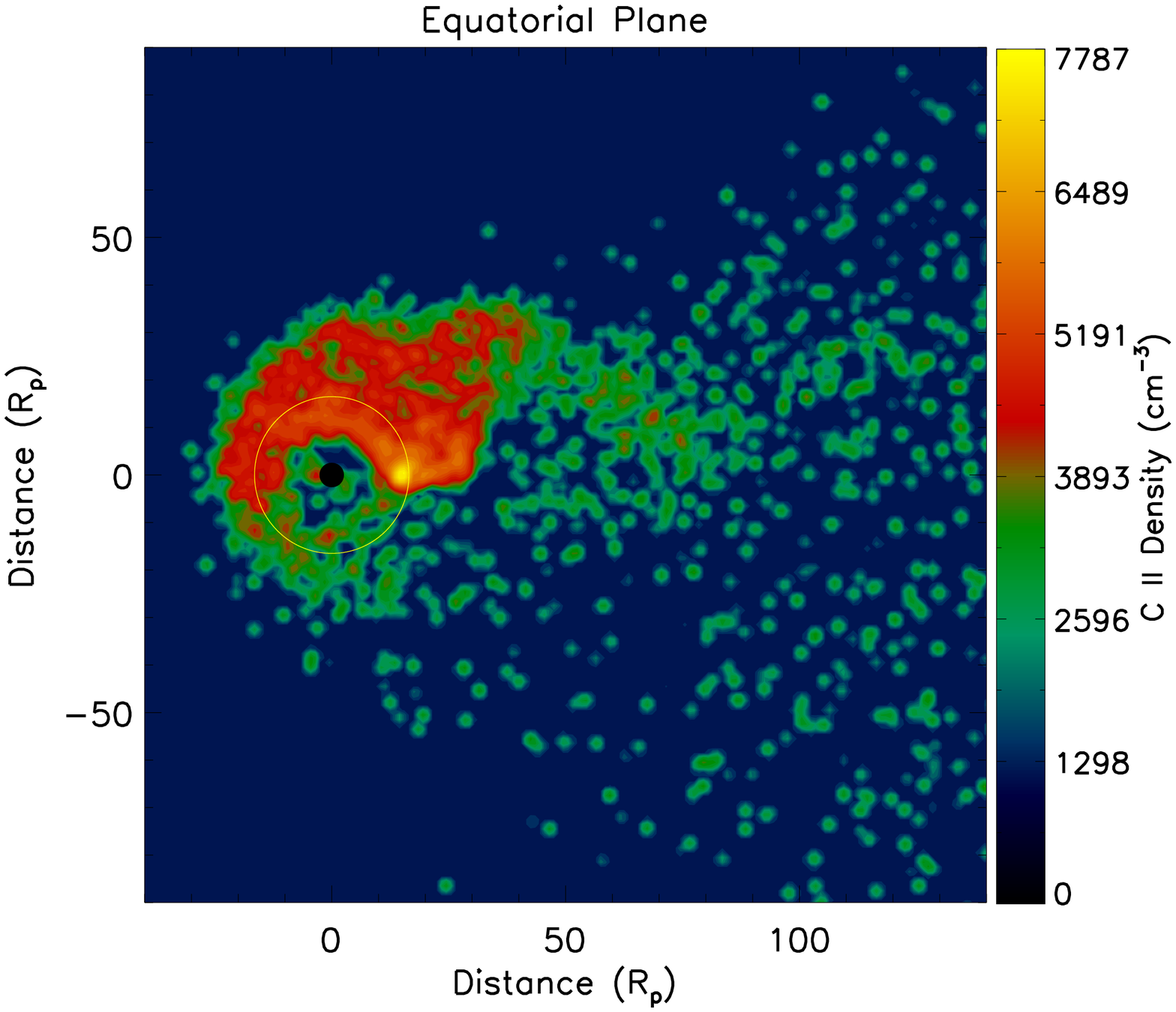}
\includegraphics[width=9cm]{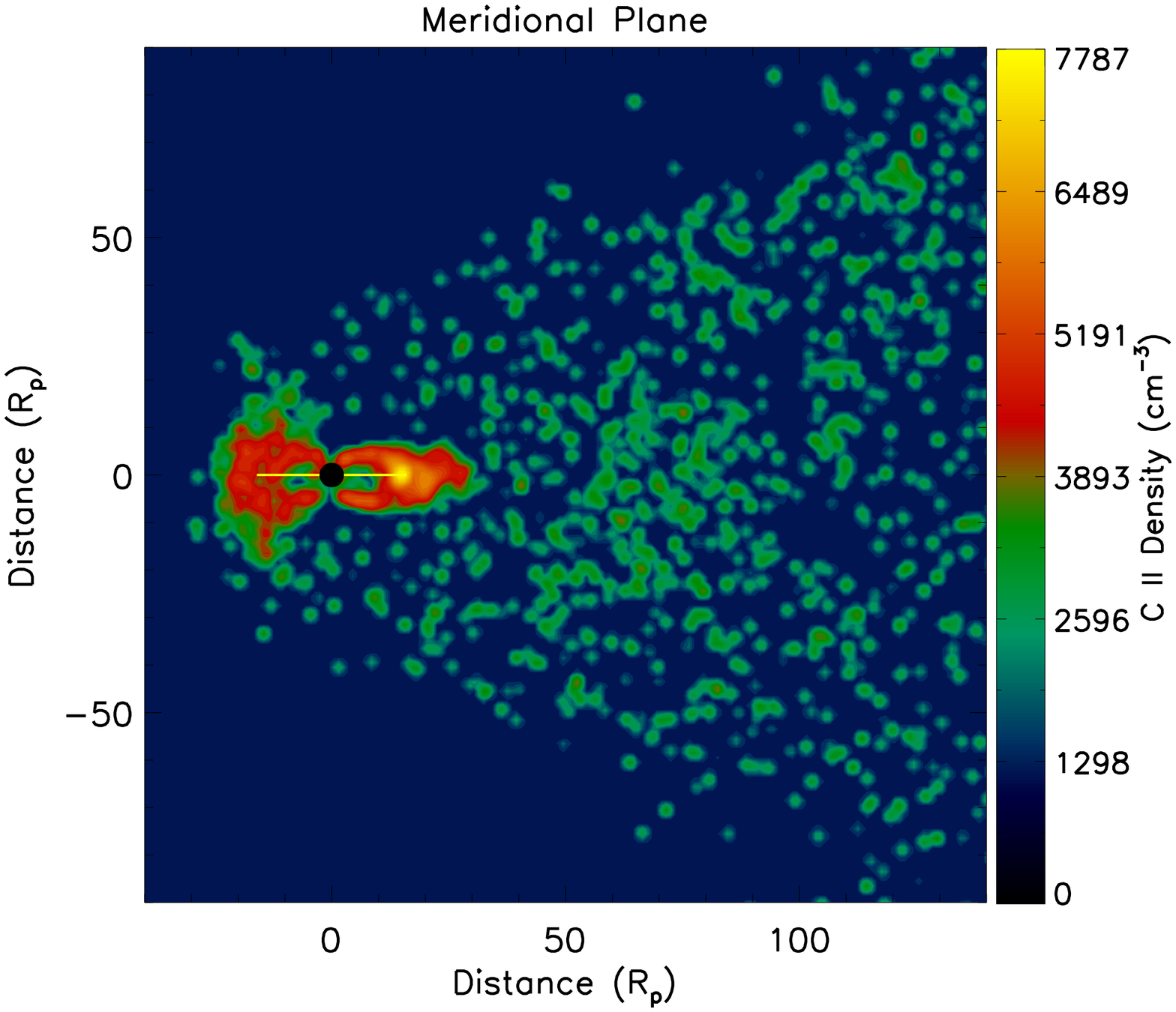}
\caption{Three-dimentional PIC simulations of exomoon plasma tori evolving within the magnetosphere cavity shown in Figure 1. Satellite orbits shown in yellow. Top left: plasma distribution of Mg\,II ions in the orbital plane for the case of WASP-12b. Extrasolar satellite orbiting at $\sim 6$\,$R_p$ from its parent planet and losing Mg\,II ion plasma at a rate of $\sim 5\times 10^5$\,g/s. Satellite at 9 o'clock. Top right: same but in the meridian plane. Bottom left: plasma distribution of C\,II ions in the orbital plane for the case of \hdn\,b. Satellite orbiting at $\sim 16$\,R$_p$ from its parent planet and losing C\,II ion plasma at a rate of $\sim 2\times10^6$\,g/s. Satellite at 3 o'clock. Bottom right: same as in left but in meridian plane.}
\end{figure*}

In the description adopted in most of the existing literature regarding exoplanetary environments, one key component has been rarely cited: the occurrence of a plasma torus that may form around an exoplanet by an orbiting moon, much like the Io torus around Jupiter \citep{kup76,bro76}. Emission lines from hot plasma tori have been indeed proposed to uncover extrasolar satellites, yet no quantitative diagnostic has been put forth  \citep{mar00}. Here, we investigate the transit absorption by a plasma torus formed by a satellite orbiting an exoplanet using a particle-in-cell (PIC) code \citep{ben13}. After a short description of our PIC model and the magnetospheric configurations of \hdn b and WASP-12b that are considered here as illustrations of the proposed new diagnostics, we will show that transit features such an early ingress absorption may appear, depending on the mass loss rate and the orbital parameters of the assumed moon.

\section{STELLAR WIND, EXOPLANET MAGNETOSPHERE, AND SATELLITE MASS OUTFLOW MODELS: THE INGREDIENTS}

The main idea of a plasma torus relies on a volcanic small satellite, like Io, orbiting a main body and loading a substantial mass of plasma to the magnetosphere of the planet. Exporting this synthetic sketch to extrasolar systems is not arbitrary, particularly for planets orbiting near their parent stars that may experience strong tidal effects in addition to the immense heating from the stellar insolation. As a component of the star-planet-satellite system, a small rocky body orbiting the planet will suffer huge tidal effects from both the planet and the nearby star. In that context, strong volcanic activity and plasma outflow are expected as predicted by several studies of volcanism on hot, rocky planets \citep{sch04,kal10}.  

Here, we use a PIC code to  simulate the plasma torus produced by hypothetical satellites orbiting exoplanets \hdn b and WASP-12b and calculate the resulting transit absorption.  As for Jupiter, the synthetic plasma torus is assumed inside the exoplanet magnetosphere. For both \hdn\, and WASP-12, we assume an isothermal stellar corona with a temperature of $1.3\times 10^6$K, corresponding to a lower limit inferred from X-ray observations \citep{for11}. {  For both systems, an interplanetary magnetic field of few milligauss is selected at the orbital position of the planet \citep{far10}. At the same orbital position, the stellar wind has a speed of $\sim 200$ km/s and a density of $\sim 7.7\times 10^2\,{\rm cm^{-3}}$ corresponding to a coronal  density of $\sim10^{10}\,{\rm cm^{-3}}$ at its base \citep{for11,lla11}. For the stellar wind parameters assumed above, the plasma around the magnetosphere obstacle is sub-Alfv\'enic \citep{coh11}. Here, we consider a planet's magnetic field strength of $\sim 7$\,G at the equator in the range thus far predicted from internal energy flux scaling \citep{chr09,rei10}. For the specific stellar and planetary parameters assumed to illustrate plasma tori formation and transit absorption, we obtain a magnetopause position of $\sim 20$\,$R_p$ (e.g., Figure 1), but we should note that other scenarios based on a hotter corona or weaker planetary magnetic field may produce a different configuration.} The planetary atmosphere is initially approximated as a spherical region of temperature $\sim 10^4$\,K and radius $\sim R_p$ that we update to correct any misfit with the observations.

\begin{figure}
\centering
\hspace*{-0.4in}
\includegraphics[width=7.5cm,angle=90]{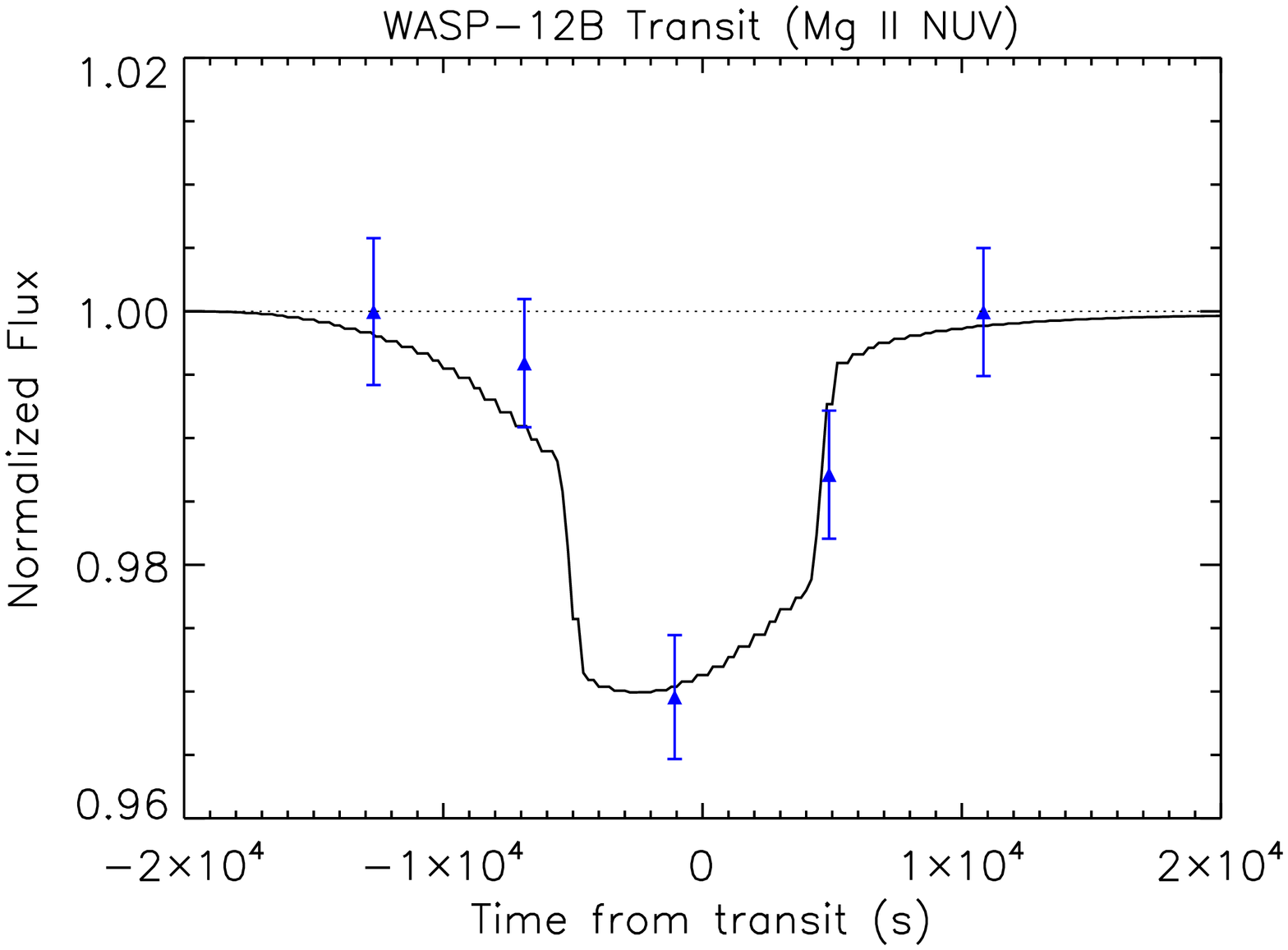}
\caption{Three-dimentional PIC simulation fit to WASP-12b Mg\,II light curve. This simulation was obtained for a moon orbiting $\sim 6$\,$R_p$ away from WASP-12b, and losing mass at a rate of $\sim 5\times 10^5$\,g/s of Mg\,II ions. This case corresponds to the plasma distribution shown in top of Figure 2. {\it Hubble Space Telescope/Cosmic Origins Spectrograph} observations are also shown for the NUV wavelength ranges described by \citet{fos10}.}
\end{figure}

The satellite is assumed a spherical body losing mass radially and isotropically from a thin $10^4$\,K envelope into the magnetosphere at a prescribed rate. The angular motion of the moon is assumed to be the planetary value because considering smaller values may dramatically increase the simulation time to achieve a single orbit of the moon\citep{hen70,she08}.

\begin{figure*}
\centering
\hspace*{-0.4in}
\includegraphics[width=7.5cm,angle=90]{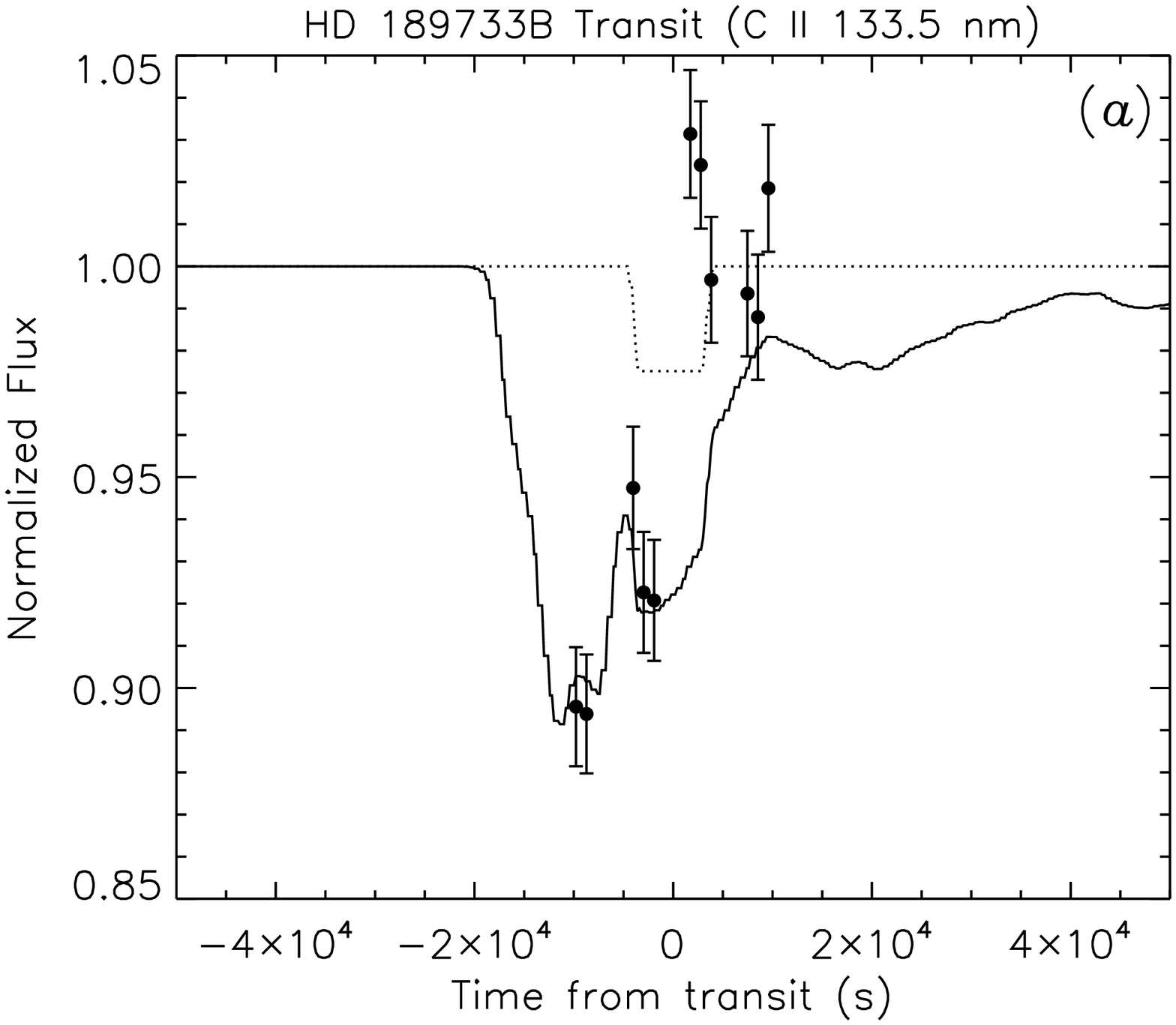}
\includegraphics[width=7.5cm,angle=90]{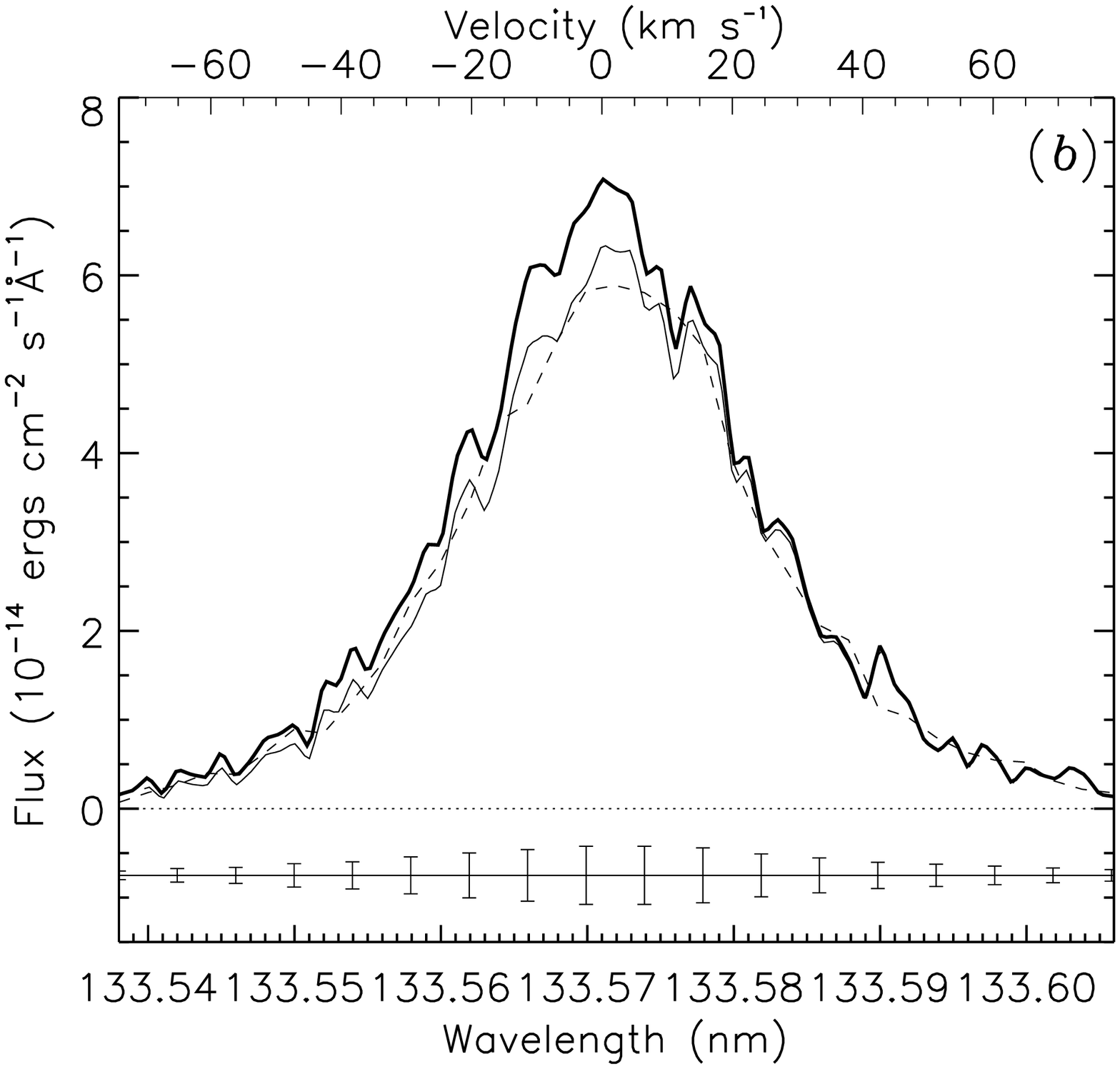}
\caption{Three-dimentional PIC simulation fit to \hdn b C\,II 133.5 nm transit. This simulation was obtained for a moon orbiting $\sim 16$\,$R_p$ away from \hdn\,b, and losing C\,II ion mass at a rate of $\sim 2\times 10^6$\,g/s. This case corresponds to the plasma distribution shown in the bottom panel of Figure 2. (a) Light curve fit. (b) C\,II line profile fit at phase position of the early ingress absorption. Comparison of model (dashed) to observations (solid) showing a fit to the transit absorption.}
\end{figure*}

\section{PIC SIMULATION OF THE PLANET-SATELLITE SYSTEM AND OPACITY CALCULATIONS}

We use an electromagnetic (EM) and relativistic PIC three-dimensional (3D) code adapted to exoplanet magnetospheres as described in detail in \citep{ben13}. The PIC code solves Maxwell's equations on a uniform grid. Charged species are sketched as macro-particles that contain a large number of real particles that may take any position inside the grid. In contrast, fields and currents are evaluated on the grid nodes. The simulation box uses a regular grid $305\times195\times195$$\Delta{r}^3$ along the three Cartesian axes. Each grid cell has a uniform size in the range $\Delta{r} = 0.2-1.0$\,\rp\ . Stellar wind is injected from the YZ plane that is orthogonal to the planet's orbit, allowing at most a number of $5.5\times 10^7$ pairs of macro-particles inside the box.  Because the PIC code is collisionless, all species experience the same fields but are kept separate in order to track their dynamics in the magnetosphere. This version of the code offers the opportunity to consider species of different charges and or masses. This key feature of the code takes its full meaning when considering three distinct sources (wind, planet, satellite) that may have different compositions and dynamics. The system is left evolve until the satellite has made few orbits. 

For the post-processing of the simulation results, plasma bulk properties are first derived as moments of the particles distribution function for each of the three sources independently. This allows us to estimate the gas opacity in the simulation box, including the local Doppler shift with respect to the line-of-sight. The transmission of the stellar disk flux through that extended volume is then derived for any orbital position of the planetary system. It is important to stress that each pixel of the projected area of the simulation box over the sky is calculated as an independent line-of-sight, which requires expensive computer resources.

\section{STELLAR FLUX TRANSMISSION THROUGH A MAGNETOSPHERE-PLANET-SATELLITE SYSTEM}
To the best of our knowledge, no prediction has ever been reported for the transit absorption of a plasma torus produced by an extrasolar satellite.
 
In a first step, we are interested in the general steady-state plasma distribution of the gas ejected by the satellite, which depends on its longitudinal position around the planet. As shown in Figure 2, the two different semi-major axes of the two moons produce different configurations (i.e., a partial versus a full torus), yet a common ribbon feature appears in the orbital plane at all times around the instant satellite position. For WASP-12b, a full torus forms along the satellite orbit that is quite similar to the Io plasma torus plasma configuration observed for several years on the Jovian system \citep{tho04}. The plasma from the satellite source fills a large volume around the planet position that contributes to the opacity usually attached to that region, yet with a majority of the density concentrated in a torus-like volume. To check the plasma configuration, we calculated the plasma density in the meridional plane (across the orbital plane) with the satellite on one edge of the magnetosphere nose-planet axis (left or right side of Figure 2). It is also  interesting to see that tubes of plasma are connecting the torus plasma north and south the orbital plane down to the planet polar regions. The plasma tubes appear denser at the elongation of the satellite body, yet the coupling is acting at most longitudes around the planet when enough plasma is available {  (however, we expect the moon-planet coupling to vanish for weak planetary magnetic field)}. As in the auroral region of giant planets of our solar system, ion and electron precipitation will induce a very efficient ionic chemistry that may profoundly modify the upper atmospheric structure, particularly at the injection latitude bands. This moon-planet EM coupling nicely reminds us of the connection between the Galilean satellites of Jupiter and the planet detected thanks to the corresponding satellite footprint of  the Jovian auroral emissions, particularly the aurora associated with the material loaded in the wake region of Io \citep{cla02}.  

For the first case of WASP-12b, we consider a moon orbiting the exoplanet with a semi-major axis of $\sim 6$\, $R_p$ corresponding to the orbital location of Io. Here, we assume that the moon produces enough Mg\,II, particularly in the proximity of the strongly ionizing XUV stellar flux. This assumption is not arbitrary in that Mg\,II was detected over a short period when a fragment of Shomaker-Levy comet was close to Jupiter \citep{fel96}. The orbit is placed inside the magnetospheric cavity and is elongated along the planet's orbit, a configuration that corresponds to a retrograde satellite captured by the planet in a later stage of its evolution \citep{hen70,she08}. Using a satellite mass loss rate of $ \sim 5 \times 10^5$\,g/s of Mg\,II ions produces a good fit to the Mg\,II light curve observed for WASP-12b in the NUV (e.g., Figure 3). With the selected semi-major axis and the moon orbit aligned perpendicularly to the star-planet line, the fitted phase for the early ingress absorption corresponds to the maximum moon elongation. However, the observations are very few which does not help locate exactly when the early ingress absorption is maximum.  The Doppler shift during the torus transit may help us disentangle the degeneracy, an information that cannot be easily used in the case of WASP-12b because of the complexity of the spectral window around the Mg\,II lines and the limited quality of available data \citep{has12}.

For our second case, we consider the C\,II early ingress absorption reported tentatively for \hdn b  \citep{ben13}.  Again, a good fit is obtained when the moon semi-major axis is close to the apparent distance corresponding to the phase position of the early ingress absorption. We thus consider a moon orbiting with a semi-major axis $\sim 16$\,$R_p$, keeping the same elongation across the star-planet line as required for retrograde satellites. In addition, we assume that the moon produces enough C\,II. The torus is only partial at 16 R$_p$ since material from the moon can only compensate for the plasma transported away in the planet's magnetotail over a fraction of the orbit. A C\,II detection from a torus around \hdn b would imply a still-volatile-rich moon which could be plausible since the system is still young with an estimated age of $\sim 1$\,Gyr. Although a full study would be necessary for a small-mass moon, this could be inferred from studies of hot Earths with outgassed atmospheres from post-formation geological activity \citep{sch12n}. The outgassing of CO$_2$ could still be significant depending on the moon's temperature, strength of geology activity and age.  As shown in Figure 2, the plasma torus has a dense region around the position of the moon, the so-called ribbon, that gives a satisfactory fit to the C\,II light curve (e.g., Figure 4(a)) if the satellite instant position crosses the projected orbit of the planet shortly before transit. At this level, it is interesting to study the spectral signature of the early ingress absorption. As shown in Figure 4(b), the blueshifted absorption that appears fits rather well with the observations. Here, it is important to stress that a larger semi-major axis may give a satisfactory fit to the C\,II light curve if we assume that the moon crosses the orbit of the planet with a different elongation. 

 At this level, it is important to stress that the simulations shown in Figures $1-4$ are specific cases selected for reasonable parameters of the stellar wind and exoplanet parameters, yet other scenarios based on different magnetospheric configurations or orbital and compositional properties of exomoons may result in different transit signatures from those reported here. In a future paper we will explore in more detail the role of the torus plasma compared to the stellar wind and planetary magnetic fields and plasmas in the dynamics and general structure of the magnetosphere. For these reasons, the cases reported here should be considered as experiments that require further investigation and more observational evidence.

\begin{figure*}
\centering
\hspace*{-0.4in}
\includegraphics[width=6.5cm,angle=90]{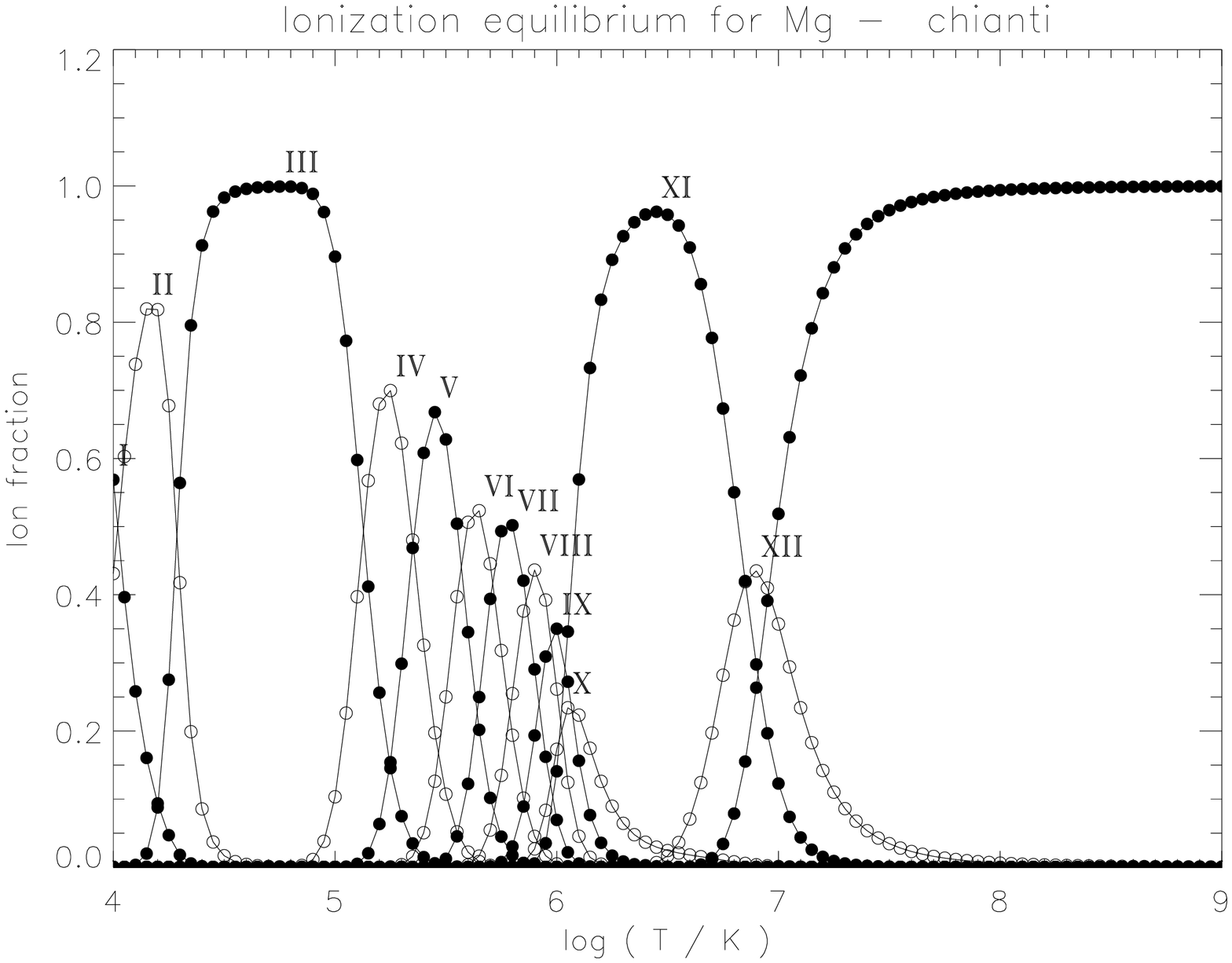}
\includegraphics[width=6.5cm,angle=90]{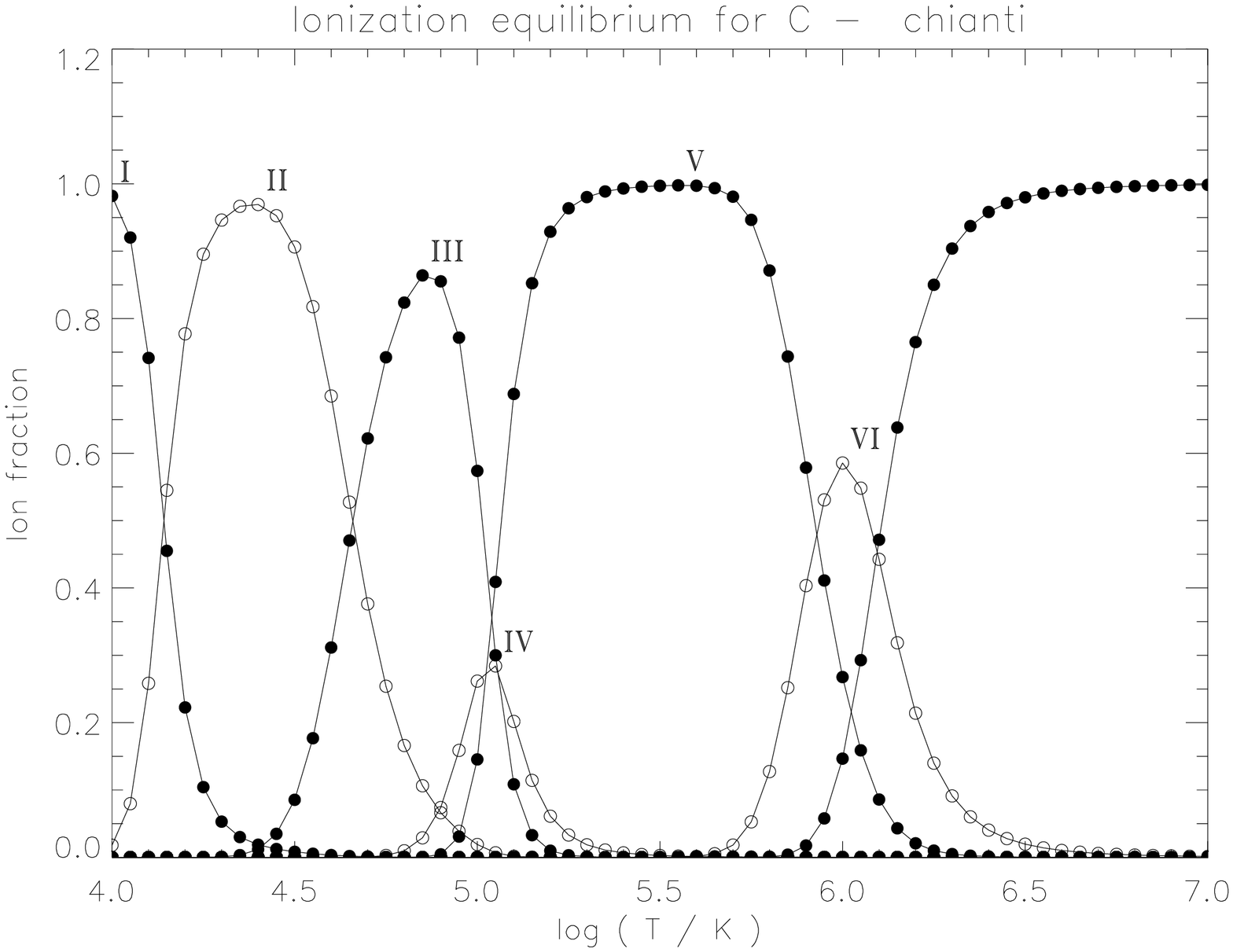}
\caption{Ionization-recombination equilibrium distribution vs. temperature calculated with the Chianti databse for Mg (left) and C (right) calculated without any interaction with other species (no charge exchange included).}
\end{figure*}

To put our preliminary results in a more general context, we tried to find whether the asymmetries predicted here have specific properties that may discriminate the moon torus signature from previous solutions advocated in the past to explain the early ingress absorption. Since 2010, it has been proposed variously that a narrow and fast stream from the planet's L1 point or shocked gas upstream of a magnetosphere or shocks resulting from the collision between planetary and stellar winds, may produce large enough absorption to be visible as an early ingress absorption \citep{lai10,vid10,kho12,fos10,ben13,bis13}. This concept was first applied to explain the early ingress absorption observed in the WASP-12b and tentatively reported for \hdn\,b. The serious problem we now see with bow-shock transit absorption is that none of the species for which the early ingress absorption was observed exists in the stellar wind or corona. Indeed, for the coronal temperature level assumed for both \hdn\, and WASP-12 ($\sim 10^6$ K), carbon and magnesium should be highly ionized as confirmed by coronal observations and in situ measurements of the solar wind \citep{sch12}. In addition, simple ionization and recombination equilibrium calculations clearly show that for the expected coronal temperature, only the highly ionized stage of heavy elements exists for $T > 10^5$\,K (e.g., Figure 5). Unless an unlikely extended and sufficiently dense planetary neutral wind (up to $\sim 6$ $R_p$ for WASP-12b and $\sim 16$\,$R_p$ for \hdn\,b) efficiently exchange charges in cascade (four to eight consecutive events) with the highly ionized stellar wind particles to produce, despite the strongly ionizing stellar radiation, enough Mg\,II or C\,II opacity upstream of the proposed shocks locally, we do not see how this explanation may stand. Also, a narrow and fast stream ($\sim 100$\,km/s) from the planet's L1 point must show a strong Doppler signature that has not been observed thus far. In that context, the early ingress absorption and other distortions of transit light curves in the UV could be the signature of ionized species that form a plasma torus around the planet. Due to the proximity of the planet-moon system to the parent star, both the strongly XUV ionizing flux and the huge tidal forces would favor the formation of such plasma tori. The occurrence of an orbiting moon around a planet that is close to its parent star is not yet well understood, yet three-body evolution models and the early-ingress UV absorptions both favor such an event to exist \citep{she08}.

\section{CONCLUSIONS} 
Here, we propose using plasma tori that are produced by volcanic moons along their orbits to detect small-size and small-mass satellites around exoplanets. Following the Io torus example, a small body with enough volcanic activity may produce a spatially extended plasma nebula that may show substantial transit absorption, particularly in the UV. The proximity to a parent star should likely enhance the volcanic activity of the small body, inherently enhancing the mass loading the planet's magnetosphere. It is therefore reasonable to posit that a strong transit signature might then be detected.

To test this scenario, we utilized our PIC 3D simulation code in which we placed a moon in orbit inside the magnetosphere cavity of a magnetized exoplanet. Without any loss of generality, we considered {\it specific} plasma conditions of exoplanet \hdn\,b and WASP-12b for illustration. For the selected cases, we show clearly that the transit absorption of a torus may be strong and varying from ingress to egress. In the example shown in Figure 4 for exoplanet \hdn\,b and C\,II line conditions, a strong early ingress absorption may appear for specific orbital parameters of the moon when assuming a $\sim 2\times 10^{6}$\,g/s outflow of C\,II ions from the small body. The total mass loss depends on the dominant species (mean mass) and the relative abundances, yet it should not exceed $\sim 2.2\times 10^{16}$\,g/year or $\sim 2\times10^{-10}$ Io$_{\rm mass}$/year. In the case of WASP-12b, a good fit to the observed Mg\,II transit light curve was obtained with a moon mass loss of $\sim 5 \times 10^{5}$\,g/s of Mg\,II ions (e.g., Fig. 3). Depending on the moon composition, the total mass loss should not exceed $\sim 1.8\times 10^{16}$\,g/year, a rate comparable to the \hdn\, value found above. Both exomoons orbital properties are consistent with the periodic and stable family {\it f} of the restricted three-body problem \citep{hen70}. 

With all the predictions proposed here, new transit observations should be obtained to definitely confirm the early ingress signature, particularly for \hdn , a bright target for which isolated ion resonances lines (like C\,II) allow a relatively easy Doppler analysis of the absorption.

\acknowledgements
L.B.J acknowledges support from CNES, Universit\'e Pierre et Marie Curie (UPMC) and CNRS in France. G.E.B ackowledges support from grant HST-GO-11576 to the University of Arizona. This work is based on observations with the NASA/ESA {\it Hubble Space Telescope}, obtained at the Space Telescope Science Institute, which is operated by AURA, Inc.
\clearpage

\end{document}